\documentclass{article}

\begin{document}
{ \Large {Gravitational attraction until relativistic equipartition of
internal and~translational kinetic energies}}

\bigskip
I.E. Bulyzhenkov

\smallskip
{Moscow Institute of Physics and Technology, Moscow, Russia}

{Lebedev Physics Institute RAS, Moscow, Russia}

%
\begin{abstract}
Translational ordering of the internal kinematic chaos provides the Special
Relativity referents for the geodesic motion of warm thermodynamical
bodies. Taking identical mathematics, relativistic physics of the low speed
transport of time-varying heat-energies differs from Newton's physics of
steady masses without internal degrees of freedom. General Relativity
predicts geodesic changes of the internal heat-energy variable under the
free gravitational fall and the geodesic turn in the radial field center.
Internal heat variations enable cyclic dynamics of decelerated falls and
accelerated takeoffs of inertial matter and its structural
self-organization. The coordinate speed of the ordered spatial motion takes
maximum under the equipartition of relativistic internal and translational
kinetic energies. Observable predictions are discussed for 
verification/falsification of the principle of equipartition as a new basic for the
ordered motion and self-organization in external fields, including
gravitational, electromagnetic, and thermal ones.\looseness=1
\end{abstract}


%
{\bf Keywords:}
{Internal energy variable,}
{GR principle of equipartition,}
{Equilibrium proximity,}
{Decelerated fall,}
{Accelerated takeoff,}
{Thermal propulsion}

\section{Introduction}
Most scientists, starting in high school, are taught the erroneous
notion that Einstein's relativistic physics in weak fields should
reproduce Newtonian physics (while not only identical mathematics) of
the point constant mass for the slow motion with $v^2/c^2 \equiv\beta
^2 \ll1$.
Despite the fact that Classical Mechanics does not know numerical
failures \lq\lq in the field of its applicability\rq\rq{,} Newtonian
mass ontology can not be true if real mechanical bodies have internal
energy. Indeed, the Einstein inertial particle has the rest energy
$M_oc^2$ (due to internal kinetic chaos), while the Newtonian mechanical body
has no internal energy content. Different degrees of freedom divorce in
principle Newton and Einstein approaches to low speed physics [1]. The
weight and inertia stay the same after warming of bodies in the Newton
theory, but not in the Einstein relativity. The goal of this paper is
to criticize the constant mass notion and argue in favor of the
internal heat-energy variable. We tend to comment quantitatively why
\lq\lq Einstein's theory can be accepted only with the recognition that
Newton's was wrong\rq\rq{} [2]. 
Today this still remains a minority view despite General Relativity
(GR) is a metric theory for chaos-order transformations and structural
self-organizations of extended matter, while Newton\rq{s} point mass
dynamics can only describe quantitative changes in the inertial system.

The equal inertial and gravitational charges in Einstein's physics are
both defined by the full relativistic energy $ M_oc^2 {\sqrt{g_ {oo}}}
/ {\sqrt{1 - \beta^2}}$ ([3], for example). This
energy counts thermal variations of the internal heat-energy of real
thermodynamical bodies. The internal heat content, $Q = Q_o {\sqrt{1 - \beta
^2}} $, and the internal temperature, $ T = T_o {\sqrt{1 - \beta^2}}$, of a
moving body obey the same law provided by a pupil of Laue [4].
This kinematic rule for heat-energy changes clarifies, for
instance, why the quantum particle in wave mechanics carries its
frequency as $ \hbar\omega= \hbar\omega_o {\sqrt{1 - \beta^2}} $.

The Laue--Mosengeil law originates, in general, from the Lorentz
transformations and their time dilation mechanism in the Minkowski
space--time. These fundamental transformations are not inherent in
Newtonian dynamics. According to an observer who calculates
relativistic energy balances, any inertial body with the rest energy $
M_o c^2 $ loses under the low speed transport some part of this
internal energy, namely $ M_o c^2 {\sqrt{1 - \beta^2}} - M_o c^2
\approx- M_o v^2/2 $. This counterflow of internal heat-energy losses
halves the kinetic energy of spatial translations $vP = M_o c^2 /
{\sqrt{1 - \beta^2}} \approx M_ov^2 \ll M_o c^2 $ at low velocities,
because $ (M_ov^2 - M_o v^2/2) = M_o v^2/2 $. 

Again, the translational
ordering of internal kinematic chaos decreases the internal heat-energy
$ {\sqrt{1 - \beta^2}}M_o c^2$. The similar energy content $k M_oc^2$
(with $1/2\leq k \leq1$) was smartly used by Nikolay Umov [4]
since 1873. According to Umov, both heat and mechanical energy can form
Newton mass flows. He considered energy perturbations in media the
primary physical process which defines the corresponding transport of
mass from known secondary relations. Contrary to the Umov vector
analysis of primary energy flows, the Newton--Euler approach to ideal
fluids is based on the scalar mass transport. Here, the balance of
energies is derived from secondary relations. But what is the most
correct way of describing the mechanical motion in reality?

The Lorentz transformations for the internal heat-energy $M_o c^2=
(E-vP)/{\sqrt{1-\beta^2}}$ and its zero 3-momentum
$(P - \beta E/c)/{\sqrt{1-\beta^2}} =0$ reveal two physical parts, $E
= M_o c^2{\sqrt{1-\beta^2}} + vP \equiv M_o c^2{\sqrt{1-\beta^2}} +
\beta^2 E $, in the relativistic structure of the summary kinetic
energy $E = M_o c^2 / {\sqrt{1-\beta^2}}$. These inseparable parts
have different kinematic origin. They can be colored under computations
like inseparable quarks in their proton. The first part $E_o{\sqrt
{1-\beta^2}}$ is the kinetic energy of internal chaotic motions or the
internal heat variable of the moving particle. The second part $M_o
c^2\beta^2/ {\sqrt{1-\beta^2}}$ is the kinetic energy of ordered
spatial translations of all chaotic densities of matter. In other
words, the Hamilton form of energy in quantum mechanics, $H = M_o
c^2{\sqrt{1-\beta^2}} + vP$, obeys the same Lorentz split over
internal and external degrees of freedom. It is not quite appropriate
to consider the formal Newton model, which mixed independent degrees of
freedom and halved the kinetic energy of ordered translations, as a
correct physical theory even \lq\lq in the field of its applicability\rq
\rq{}. Newton's physics with only one degree of freedom for two (chaos
and order) competing kinds of kinetic energy is unable to provide
reliable referents for the relativistic motion and self-organization of
real matter with the internal heat variable.\looseness=1
\section{Inertia depends on heat contrary to~Newton}
In general, it is necessary to recognize the model defectiveness of
Classical Mechanics with the point mass concept. Contrary to Newton,
there is no need at all to trace the conservation and transport of
scalar masses in the Einstein theory. The latter incorporates
thermodynamics though the common balance of heat and mechanical energies.
The time-dilation effect for moving internal energy is well tested due
to multi-kilometer tracks of cosmic muons in the atmosphere.
Because of the suitable notion of the internal heat variable, $Q (v^2)
= Q_o {\sqrt{1-\beta^2}} \neq \mathrm{const}$, one could completely abandon the
inertial mass notion $M_o=\mathrm{const}$. The constant point mass is
independent on variations of the body temperature and the body speed~$v$. But mechanical inertia is an energy characteristic since it grows
with the absorption of energy of \lq\lq massless electromagnetic
waves\rq\rq{}. The traditional preservation of two competing notions
(rest energy and mass) for the inertial charge of a real thermodynamic
body inevitably leads to physical confusion at first, and then to
incorrect mathematical models. Below we demonstrate the inability of
the Newtonian theory of \lq\lq cold\rq\rq{} point masses to act as a
true limit for Einstein's General Relativity of inertial energies with
the internal heat variable.\looseness=1

We define the full energy of an isolated (or remote from other field
centers) particle at rest $M_oc^2 \equiv Q_o$ as its internal
heat-energy $Q_o$ associated with the chaotic motion of elementary
material densities. The latter can be Ricci metric densities of a
continuously extended particle in a Cartesian material space or can be
particle\rq{s} internal sub-elements (like point quarks) in the empty
space alternative of Newton. The internal chaotic motion can correspond
to Descartes mechanical vortexes which are shaping particular spatial
structures of elementary matter. We admit an equilibrium internal state
behind the heat-energy variable $Q (v^2) = Q_o {\sqrt{1-\beta^2}}$
despite we associate this kinetic energy with internal relativistic
heat or internal chaos. And this internal heat is independent from
external fields. The full rest energy at the finite distance R from the
center of the radial gravitational field, $E_o(R) = Q_o {\sqrt
{g_{oo}(R) }}$, is defined by both the internal heat $Q_o$ and the
metric field contribution from $g_{oo}(r=R) = \mathrm{const} < 1 =
g_{oo}(r=\infty)$.
The free gravitational fall in a constant radial field from the
coordinate R, where $v^2(r=R) = 0$, will keep the initial relativistic energy
$E (r,t)\equiv{ Q_o \sqrt{g_ {oo}(r,t) } }/ {\sqrt{1- \beta^2(r,t)}}
= E_o(R)$ [3]. We decompose this probe body
energy into three path-dependent parts
%
\begin{eqnarray}\label{eq1}
E (r,t) &\equiv&\frac{ Q_o \sqrt{g_ {oo} (r,t)} } {\sqrt{1- \beta^2(r,t)}} \equiv{Q_o} {\sqrt{{1-
\beta^2(r,t)}}}
\nonumber
\\
&&{}+ \biggl[\frac{Q_o \beta^ 2(r,t)} {\sqrt{1- \beta^ 2(r,t)}} - \frac{ Q_o (1 - \sqrt{g_ {oo} (r,t)} )} {\sqrt{1 - \beta^ 2(r,t)}} \biggr].
\end{eqnarray}
This full relativistic energy is proportional to the value of the total
inertial (equal to gravitational) charge $ q(r,t) \equiv E (r,t) /
\varphi_o$ of the falling body (where $\varphi_o \equiv c ^ 2 / {\sqrt
G} = 1.04 \times 10^{27} $ eB). Despite the full charge $q$ is
constant in the constant gravitational field, we will study (in the
absence of rotations) its path dependent contributions from the
internal heat-energy $Q(r[t]) \equiv{Q_o} { \sqrt{1- \beta^
2(r[t])}}> 0 $ of the relativistic body, from the kinetic energy of
ordered spatial translations \mbox{$ T(r[t]) \equiv{v}{P} ={Q_o\beta^
2(r[t])} / {\sqrt{1- \beta^ 2(r[t])}}> 0 $} as a whole system, and
from the negative potential shift $ U(r[t]) \equiv  - Q_o [1- {\sqrt
{g_ {oo}(r[t])}} ] /
{\sqrt{1- \beta^ 2(r[t])}} \leq0 $ in the external
gravitational field, with $ 0 \leq g_ {oo}(r[t]) \leq g_ {oo}(R)$ and
$r[t]\leq R$. Ernst Mach would probably approve our definition of
charges, $ q \equiv E/\varphi_o$, based on their full energy. Indeed,
body\rq{s} inertia and gravitation depend on the distribution of all
remote bodies through gravitational contributions to a purely
mechanical part of the moving body energy,
$ M(r,t) \equiv T(r,t) + U(r,t) = E(r,t) - Q(r,t) = Q_o [\beta^
2(r,t) - 1 + {\sqrt{g_ {oo}(r,t)}}] / {\sqrt{1- \beta^ 2(r,t)}}$.
This mechanical part $ M(r,t)$ is grouped in (1) inside the square brackets.

The summary kinetic energy of internal (chaotic) and translational
(ordered) motions, $ K(r,t) \equiv Q(r,t) + T(r,t) = {Q_o} / {\sqrt{1-
\beta^ 2(r,t)}} $, can be also traced in (1). This combination was
implemented in Special Relativity only in favor of old referents of
Newtonian dynamics. In other words, Special Relativity in its
nonrelativistic limit was oriented to the Newtonian kinetic term $Q_o
\beta^ 2/2 $ for the net transport of internal and external energies.
Contrary to its capabilities, Special Relativity did not recognize
internal energy changes in moving thermodynamical bodies. Rapid path
variations of the internal heat-energy
$ {Q_o} {\sqrt{{1- \beta^ 2}}} $ reveal collinear counter changes of
thermal (chaotic) and translation (ordered) energy flows of the point particle.
The model dynamics of Newton's cold masses should in no way serve as a
non-relativistic referent for a continuous relativistic medium with
heat-energy flows. These vectors in liquids and gases may have
non-collinear options regarding to ordered translations of material densities.
Instead of the Lorentz vector transport of two independent and
competing energy flows, the Newton--Euler transport of the summary
density amplitude (of these non-collinear vectors in continuous media)
is declared nonphysically for the (constant) mass transfer along
ordered translations, but never along heat exchanges.

\section{Metric self-sufficiency of General Relativity with charges of~heat-energy}\label{sec3}

The relevance of Newton's theory should be revised not only for the
slow dynamics of inertial energies with heat in Special Relativity, but
also for the search of metric solutions of General Relativity through
the Newtonian limit of weak fields. Indeed, the inhomogeneous metric
component $ g_ {oo} (x) $ can be uniquely expressed from the most
general balance $E(x) \equiv Q_o {\sqrt{g_{oo}(x) }} /{\sqrt{{1-
\beta^ 2(x)}}} $ and the universal ratio $ U(x) / E(x) \leq0$ of the
negative gravitational contribution $U(x)$ to the full energy content $
E(x) \equiv T(x)+U(x)$:
%
\begin{eqnarray}
{\sqrt{g_{oo}(x)}}&\equiv&\bigl[T(x)+U(x)\bigr] \frac {\sqrt{1-\beta^2(x)}} {Q_o}
\nonumber \\
&\equiv& 1 +\frac{U(x) {\sqrt{g_{oo}(x)}} }{E(x)} \equiv \frac{1 }{ [1 +(-U(x)/E(x))] }.\nonumber \\
\label{eq2}
\end{eqnarray}

On the basis of algebraic identities in relations (2), one can
formulate the following $g_ {oo}$-theorem: \lq\lq The temporal metric
component is defined by the gravitational potential
$\varphi(x) \equiv U(x) / E(x)$, with $-\infty< \varphi(x) \leq0$,
strictly as $ {g_ {oo}(x)} = 1/[1 - \varphi(x)]^2 \leq1$\rq\rq{.}
The static field potential $ U (x) /E (x) = -\mathit{GE}_2 / c ^ 4 r $ of the
large energy integral $E_2 =\mathrm{const} \gg E(x)$ matches the Newton
gravitational potential in the weak field limit and corresponds to
peculiarity free metric component $g_ {oo} (r) = [c ^ 4r / (c ^ 4r +
\mathit{GE}_2)]^2$ in strong fields. Therefore, the empty space Schwarzschild
metric [6] 
with $ g_ {oo} (r) = (c^2r-2\mathit{GM}_2) / c^2r $ for strong fields does not
agree with (2) and with the Einstein--Infeld material fields
[7,8] of inertial energies with
internal heat.
It is unjustifiable to equate the model gravitation of Newton's cold
and point masses in empty space to the Mach--Einstein gravitation of
heat-energy variables in a Cartesian non-empty space.


Recall that in 1913 Einstein and Grossmann introduced a gravitational
field only into the time section of the Minkowski space--time interval.
They also preserved, without variants at that time, the point mass
model or localized substance for comparison of relativistic and
classical motions in gravitational fields [9]. 
After the November 1915 correspondence with Gilbert, Einstein had a
tensor equation for metric gravitational fields. The space--time
interval $ ds = {\sqrt{d \tau^ 2 - dl ^ 2}} $ became a combination of
two warped sub-intervals, $ d \tau^ 2 \equiv{{g_ {oo}}} [cdt + (g_
{oi} dx ^ i /g_ {oo})] ^ 2 $ and
$ dl ^ 2 \equiv(g_ {oi} g_ {oj} g ^ {- 1/2} _ {oo} - g_ {ij}) dx ^
idx ^ j $. In this case, the non-geometrized substance in the right
hand side of the Einstein Equation [10] 
again allowed spatial localization and empty space regions in
accordance with Newton's referential presentations of localized
inertial masses. But in 1938, Einstein criticized the duality of
massive particles and massless fields. He pointed to the non-dual way
in the relativistic physics of continuous carriers of energy
[7].
After that, the author of General Relativity clearly characterized the
Schwarzschild 4-interval with a singularity as \lq\lq not related to
physical reality\rq\rq{}\ [11]. But where is the invisible
revolution of new Einstein's ideas of 1938--1939 after the first shocks of
Newtonian gravity in 1913--1916?

\section{The non-dual turn of 1938 from the dual gravitation of 1916}

The Schwarzschild metric for curved emptiness was rejected for physical
reality not only by Einstein. A~curved 3-space in the metric solution
of 1916 for a constant central field [6] 
could not suit many opponents. Sommerfeld, Schwinger, Feynman and many
other physicists sought to preserve the Euclidean nature of
electrodynamics, including its constant Gaussian flux through a
two-dimensional closed surface and the strict Bohr--Sommerfeld
quantization rule over a closed line contour. In the 1980s, Russian
academician A.~Logunov
suggested the alternative theory of gravitation
[12] 
by returning back to the Minkowski space--time of 1908. This would
return flat space to relativistic physics. According to Logunov, it
would eliminate nonphysical black holes in strong fields and would lead
to a cyclical state of matter with high and low densities.

Logunov's program goals should be welcomed, especially since Einstein
himself tried to improve the relativistic theory. However, Einstein
understood that consistent classical mechanics should be constructed,
like quantum mechanics of the spatially distributed particle, in
non-dual terms of extended energy densities: \lq\lq Matter is where the concentration of energy is great, field is where
the concentration of energy is small. But if this is the case, then the
difference between matter and field is a quantitative rather than a
qualitative one. There is no sense in regarding matter and field as two
qualities quite different from each other. We cannot imagine a definite
surface separating distinctly field and matter.\rq\rq

The non-dual theory of energy densities can not be transformed into a
dual one because of the transition from high concentrations of matter
to low.
It is unreasonable to expect that non-dual objects in physics of the
microworld will begin to change into dual ones due to scaling of space
or speed values. The model separation of the continuous energy
macrocosm into an allegedly localized substance within weak fields can
be explained by the local quantitative observations of nonlocal
reality. But any observations can not be justified as true solely with
the means of a mathematically correct theory.
In 1938, Einstein proposed not to return back to the sum of three main
concepts, space--time plus field plus matter, as Logunov chose [12] 
instead of two main concepts of 1916 (space--time-field plus substance).
Contrary to Logunov, Einstein proposed to go ahead and
to diminish the number of main physical concepts to one. He
qualitatively integrated the continuous density of the extended
elementary particle into a united space--time-field-substance. This
revolutionary proposal was not accompanied at that time by new formulas
and verifiable predictions. But Einstein immediately warned from the
non-dual theory of pure fields that a continuous particle must
disappear from the auxiliary right-hand side of his 1915 tensor
equation. Today everyone can check that the self-sufficient GR interval
with $ g_ {oo} (r) = r ^ 2 / (r + r_o) ^ 2 $ admits 6 inherent metric
symmetries for the Euclidean spatial section, $ dl ^ 2 = \delta_ {ij}
dx ^ idx ^ j $, and for the vanishing Einstein curvature $ G_ {oo} = 0
$ of the static material field. The static curvature nullification
corresponds to the static Einstein Equation, $R_ {oo} = g_{oo} R/2 \neq
0 $ [1]. The extended particle belongs to the field part
of the Einstein Equation rather than to its right hand side.

\section{Turning traffic in the static gravitational center}

To predict new testable effects in the Descartes pa\-ra\-digm of the
nonempty space of extended energy-charges, it is necessary to firmly
reject the New\-to\-nian referents from the point mass physics. It is
suf\-fi\-cient to assume that the relativistic rest energy $ E_o $ has
the pure kinetic nature due to internal motions of metric material
densities within the extended (non-point) elementary particle. Then,
the remote probe body at rest will have in (1) a maximum of its
internal (kinetic) heat-energy $ E (\infty) = Q_o $ under zero
mechanical energy: $ M (\infty) = T (\infty) + U (\infty) = 0$, with $
T (\infty) = U (\infty) = 0$.
At any final removal of $ R $ from the static gravitational center, the
probe body at rest (when $ \beta(R) = 0 $ and $ T (R) = 0 $) possesses
the diminished full energy $Q_o {\sqrt{g_ {oo} (R)}}$ but the same
internal heat-energy $Q_o(r) = Q_o = \mathrm{const}$. The point is that the body
\lq gained\rq{} a negative potential energy $ M (R) = U (R) = -Q_o
[{\sqrt{g_ {oo} (\infty)}} - {\sqrt{g_ {oo} (R)}}] < 0$.

The initial rest energy of the thermodynamic probe body,
$Q_o + M(R) = Q_o {\sqrt{g_ {oo} (R)}} \equiv E_o(R) > 0 $, stays
positive in strong gravitational fields apart from their center, \mbox{$g_
{oo} (0) = 0$}.
Without the external gravitational potential or other external
stimulation, the chaotic equilibrium state of probe matter does not
have energy reasons for further spatial self-ordering. Indeed, the
positive rate of translational energy changes under spatial
self-accelerations cannot be balanced in (1) by the negative rate of
internal heat changes only, $\delta[ K(v^2) + Q(v^2)] \neq0 $.
For the ordered spatial motion with the growth of~$v^2$, interacting
mechanical bodies have to create mutual gravitational potentials and
relevant conditions for mutual rotations.

Let us return to the free fall from a fixed height $ R $ on a heavy
static center with a large energy $ E_2 \gg E_o (R) $. The constant
field of this gravitational center keeps the initial energy $ E_o = Q_o
{\sqrt{g_ {oo} (R)}} $ of the falling thermodynamical body, $ E (r[t])
= Q_o
{\sqrt{g_ {oo} (r) / [1- \beta^ 2 (r)]}} = Q_o{\sqrt{g_ {oo} (R)}}
$ for $0\leq r[t]\leq R$. Such a GR energy conservation allows
to associate $g_ {oo}(r)$ with the relativistic physical speed $v(r)
\equiv c d r/d\tau= dv/dt {\sqrt{g_ {oo} (r)}}$, where $ 0 \leq v ^ 2
(r) < c ^ 2 $, as well as with the coordinate speed $dr/dt $. The
latter can be measured in practice by a remote observer with the
available clock rate $dt \equiv dx ^ o / c$. Again, we start
computations from the known GR equation, $Q_o{\sqrt{g_ {oo} (r)}}/
{\sqrt{1-\beta^2(r)}} = Q_o{\sqrt{g_ {oo} (R)} } $, for the
relativistic energy conservation under radial incidence of a probe
charge-energy $Q_o/\varphi_o$ in a constant radial field. This equation
reads for the coordinate speed $dr/dt$ as
to associate $g_ {oo}(r)$ with the relativistic physical speed $v(r)
\equiv c dr/d\tau= dv/dt {\sqrt{g_ {oo} (r)}}$, where $ 0 \leq v ^ 2
(r) < c ^ 2 $, as well as with the coordinate speed $dr/dt $. The
latter can be measured in practice by a remote observer with the
available clock rate $dt \equiv dx ^ o / c$. Again, we start
computations from the known GR equation, $Q_o{\sqrt{g_ {oo} (r)}}/
{\sqrt{1-\beta^2(r)}} = Q_o{\sqrt{g_ {oo} (R)} } $, for the
relativistic energy conservation under radial incidence of a probe
charge-energy $Q_o/\varphi_o$ in a constant radial field. This equation
reads for the coordinate speed $dr/dt$ as
%
\begin{equation}\label{eq3}
\frac{ g_{oo}(r)} {{g_{oo}(R)}} = 1- \frac{v^2(r)}{c^2} = 1 - \frac
{1}{ g_{oo}(r)c^2} \biggl(
\frac{dr}{ dt} \biggr)^2.
\end{equation}

Taking into account both directions of the radial motion, the second
equality in the strong field equation (3) can be rewritten for the
coordinate velocity in an equivalent form with the introduction of a
unit vector ${\hat{\mathbf{r}}} \equiv{\mathbf{r}}/r$:
%
\begin{equation}\label{eq4}
\frac{d{\mathbf{r}}(t)}{dt} =\pm{\hat{\mathbf{r}}} c \sqrt{g_{oo}(r) \biggl[1-
\frac{g_{oo}(r)} {g_{oo}(R)} \biggr]}.
\end{equation}
It turns out from this universal relation of General Relativity that
from the viewpoint of the remote observer the body first dials the
coordinate speed, $ \mathbf{0} \rightarrow d \mathbf{r} / dt \rightarrow-
{\hat{\mathbf{r}}} c {\sqrt{g_ {oo} (R)}} / 2 $, in the region of the
moderate fields $ {{g_ {oo} (R)}} \geq g_ {oo} (r \rightarrow r_ {eq} )
\rightarrow {{g_ {oo} (R)}} / 2 $. Then, with further growth of the
field as the probe body moves toward the center in the region $ {{g_
{oo} (R)}} / 2 \geq g_ {oo} (r \rightarrow0) \rightarrow 0 $, this
body begins to decelerate to a zero coordinate speed,
$ - {\hat{\mathbf{r}}} c {\sqrt{g_ {oo} (R)}} / 2  \rightarrow d {\mathbf{r}} / dt \rightarrow{\mathbf{0}} $.
In this case, the magnitude of the
physical speed $ | {\mathbf{v}} | = | d {\mathbf{r}} / {\sqrt{g_ {oo} (r)}} dt |
$ is growing continuously from the equilibrium physical speed $ v(eq) =
c/{\sqrt2}$ to the speed of light limit $c$ next to the center of the
radial field. The body completely loses its chaotic
heat-energy at the end of the fall, $Q(r=0,\ v^2=c^2) = 0$. This lack of
the internal energy allows the physical velocity of the inertia-free
particle to formally reach the speed of light in the very center of
gravity. In contrast to the growing physical speed $ v(r \rightarrow0)
= dr/{\sqrt{g_{oo}(r)}}dt \rightarrow c $, the coordinate speed
diminishes to zero at the center, $ dr (r \rightarrow0)/ dt
\rightarrow0 $, where the probe body with completely ordered kinetic
energy temporarily stops before the start of its reverse takeoff.

After an unstable stop in the extreme nonequilibrium state with $r=0$, $Q
(0) = 0$, $M (0) = K (0) + U (0) = Q_o {\sqrt{g_ {oo} (R)}} = E_o$ in
the very center of the radial gravitational field, the probe body
starts the reverse accelerated motion toward the equilibrium radial
proximity at $r_{eq}(R)< R$. After passing through the equilibrium
proximity from the center, $ r [t] = r_{eq}(R)$, the probe body
continues to increase its vertical height and the internal chaotic
energy $Q (v^2[r,t])$. Ultimately, the body returns with Newtonian
deceleration to the height~$ R $, where it stops in the initial
nonequilibrium state with the same full energy $E_o = Q_o {\sqrt{g_
{oo} (R)}}$,
$ Q = Q_o$, $K (R) = 0$, and $U (R) = - Q_o [1 - {\sqrt{g_ {oo} (R)}}]$.

In order to find the observable (coordinate) acceleration of the probe
body on the entire trajectory of the free gravitational fall and the
reverse takeoff, it is sufficient to differentiate the coordinate
velocity (4) over the world (coordinate) time
$ t=x^o/c $ and to use the regular computation rule $dg_{oo}(r[t])/dt
= (dg_{oo}/dr) dr[t]/dt $:
%
\begin{eqnarray}
\frac{d^2 \mathbf{r}[t]}{dt^2} &=&\hat{\mathbf{r}}c^2 \biggl( \frac{1}{2} -
\frac{g_{oo}(r)}{g_{oo}(R)} \biggr) \frac{dg_{oo}(r) } {dr} \label{eq5}
\\
&\Rightarrow &
\cases {
-(1- \frac{4r_o} {r} ) \frac { c^2 r_o {\mathbf{r}} } {r^3}, \ \nonumber
g^{1916}_{oo} = 1 - \frac{2r_o} {r}
\cr 
- \frac {c^2r_o {\mathbf{r}}(r^2-2r_or-r_o^2) } {(r+r_o)^5 }, \ 
g^{2008}_{oo} = \frac{r^2} {(r+r_o)^2} \cr \nonumber 
}.   \ \ \ \ \ \ \ \ \ \ \ \ \ \ \ \ \ \ \ \ \ \ \ \ \ \ \ \ \   (6) 
\end{eqnarray}

We considered the gravitational fall from the infinity in (6), where $
R \Rightarrow\infty$, $ g_ {oo} (\infty) = 1$, and $ g_ {oo} (r_
{eq}) = 1/2 $. And we employed in (6) both the Schwarzschild metric of
1916 for the empty space around a point source and the metric solution
$ g_ {oo} = r ^ 2 / (r + r_o) ^ 2 $ from the algebraic identity (2)
with $ U / E = -r_o / r $ and $ r_o \equiv \mathit{GE}_2 / c ^ 4 $ [1].
 In both cases, the Newtonian acceleration by the weak field, $
- c ^ 2 r_o {\mathbf{r}} / {r ^ 3} $, smoothly changes the sign for the
deceleration at $ r> 2r_o $, i.e. above the supposed black hole horizon
in conventional dual physics with empty space.

The equilibrium radial level $r_{eq}(R)$ (at which the coordinate
acceleration (5) vanishes for the most general case of metric relations
${g_{oo}(r_{eq})} = {g_{oo}(R)}/2$) depends on the initial height $R$
of the radial fall. However, the physical speed parameter $ \beta
^2_{eq} = 1/2$ is universal for all equilibrium levels $r_{eq}(R)$ as
can be found from the GR equations (3). The derived relation $v^2_{eq}
= c^2/2$ for each and all particles on cyclic geodesic falls-takeoffs
maintains the exact equipartition of disordered (internal) and ordered
(translational) kinetic energies,
%
\begin{equation}\label{eq7}
Q (r_ {eq}) \equiv{Q_o} {\sqrt{{1-
\beta^ 2 _ {eq} }}} = \frac{Q_o \beta^ 2 _ {eq} } {\sqrt{1- \beta^ 2 _ {eq}} } 
\equiv T
(r_ {eq}),
\end{equation}
at all zero acceleration spheres of radii $r_{eq}(R)$ around the strong
field center.

The Newton gravitational fall turns out to be in General Relativity a
universal tendency of a rest energy particle toward the equipartition
of its disordered and ordered kinetic energies. This universal tendency
of each elementary particle to distribute the relativistic kinetic
energy equally over internal and external degrees of freedom can be
formulated as the GR principle of equipartition. It stands behind the
universal capacity of elementary matter to reorganize its internal
heat-energy in the environment of external fields. Recall that there
are no internal degrees of freedom in Newton's physics which can employ
gravitation only for quantitative changes of mechanical values rather
than for qualitative creations of new spatial structures of energies
due to chaos-order transformations. Newtonian mass without the internal
heat variable would never return back from the strong field region.
There are no clear options for the structural evolution of matter in
Newtonian reductionism with degenerated energy degrees of freedom.
Therefore, Newton's dynamics maintains only the growth of entropy up to
the thermal death. General Relativity maintains both dynamical and
structural changes in the chaos vs order balances. The new kinematic
tendency of the internal heat-energy to the dynamical equipartition
with the ordered kinetic energy allows to accept Einstein's theory
\lq\lq with the recognition that Newton's (gravitation and mechanics) was wrong\rq\rq{}
[2].
\section{Conclusions on verifiable predictions}
The internal heat variable is important to take into account in
mechanics of liquids and gases even for non-relativistic laboratory flows.
Contrary to the collinear transport of internal and translational
energies in one elementary particle, heat transfer in multiparticle
media can not always be directed along the vector of ordered
translations. Despite Umov\rq{s} vector  transport and join conservation 
of heat and translational flows of energies in tensor thermomechanics  [5], 
Newtonian proponents of the scalar mass
transfer still maintain the homogeneous pressure in the radial section
of the fully developed laminar flow of liquid in rectilinear pipes
[13]. It is not surprising that they register
often unexpected (for them) extra heat and pressure on side boundaries
of hydrodynamical energy devices like Ranque--Hilsch vortex tubes,  De
Laval and Francis turbines, jet engines, etc. But it is surprising that they do not associate
turbulence with the kinematic tendency of media to new forms of equilibrium structural
organizations under external impacts.

So far we neglected, for simplicity, all wave exchanges of a probe
thermodynamic body with third bodies (or with the world thermostat)
under the considered free fall with the Laue--Mosengeil cooling of
initial thermal energy. But in such an adiabatic approximation, the
geodesic decrease in the internal energy variable (due to the increase
in the translation energy) must be accompanied by a decrease in
temperature and a change in the Stefan--Boltzmann radiation from the
cooling body under consideration. Therefore, the interpretation of
gravitational events through the internal energy variable can suggest a
kinematic self-cooling/self-warming mechanism for geodesically moving
asteroids or comets. This thermal geodesic mechanism predicts, for
example, rapid heat changes and volcanic activities for small
satellites of large planets due to sharp decelerations in the inertial
system of distant stars (related nowadays to the cosmic microwave background).

If the kinematic tendency to the energy equipartition between chaos and
order is the fundamental property of the mechanical motion, then this
tendency should stay behind all spatial accelerations toward new
equilibrium states under any renovations of external fields. Any
constant dimensionless field $\Theta({r})$, which can change in (1)
the initial rest energy, $Q_o\Theta(R) \neq Q_o$, like in $E_o(R)
\equiv Q_o {\sqrt{ g_{oo}(R) } }$ for gravitation, could result in a
propulsion similar to an accelerated motion in the metric field
$g_{oo}(r,t)$. Such an external field can be modeled both from the
electric Coulomb potential and from heat exchangers with distributed
external sources/absorbers of thermal energy. The external thermal
potential can deliver its contribution to the internal energy of the
probe body quite rapidly under ideal (very fast) exchanges. Any
inertial body at rest tends to drop excess of its internal thermal
energy to low temperature bodies and tends to avoid thermal energy from
high temperature bodies. In a relevant thought experiment, a small
probe body at rest should start to move toward lower thermal potentials
like in the case of lower gravitational potentials. For example, the
bullet in the barrel without friction should move toward the cold end in tensor thermomechanics for Umov's energy flows.
Such kind of experiments, say on the International Space Station, can
maintain or falsify the GR principle of kinetic energy equipartition
behind the free spatial motion of matter and its structural
self-organizations. The thermal propulsion by  the warmed material space in question could be
also studied for moving bodies on almost frictionless equipotential
surfaces with hot (sunny) and cold (shadow) areas. Tensor thermomechanics of non-empty space might
assist to find proper energy approaches to \lq the sailing stones\rq\
at Racetrack Playa, California.

The point mass physics is to be replaced by internal energy physics in
many relativistic phenomena as well. The Einstein gravitational theory
of 1916 with Newtonian referents for the Schwarzschild metric of strong
static fields predicts the presence of the gravitational horizon at
distances $ r = 2r_o \equiv2\mathit{GE}_2 / c ^ 4 $ before the center of the
gravitational charge $E_2/\varphi_o$. However, the similar metric
solution of General Relativity in the 1938 Einstein---Infeld paradigm
of material fields predicts the smooth decrease of $g_ {oo}(r)$ and the
observable radial fall/takeoff of probe bodies with internal heat up
to $r = 0$. Confirmation or refutation of the black hole horizon in the
center of the galaxy can refute or confirm, respectively, the 1938 idea
of Einstein and Infeld to redesign the world space continuum in
non-dual terms of material fields without the localized particle. The
ontological merger of extended substance and strong energy fields can
realize the criterion of double unification known since the middle of
the last century: to unify a particle with a field and electricity with
gravity [8]. \looseness=1 

The predicted deceleration of vertically falling matter with further
takeoff acceleration, as if along a degenerated Keplerian orbit, could,
upon detection, clarify the kinematic role of the internal heat
variable. The new approach to gravity through the principle of
equipartition of relativistic kinetic energies could be examined in
relevant cosmic observations. By taking $r\ll r_o$ in the relativistic
generalization (6) of the Newton gravitational law one can simulate the
accelerated expansion of the Metagalaxy based on the previous cyclic
compression with deceleration.

The radial motion to the center of the galaxy from its periphery is
characterized by the kinematic cooling of thermodynamic bodies and,
therefore, by the non-equilibrium absorption of radiation energy from
the world thermostat. This implies the presence of cold (dark) regions
for IR and optical observations.
Inelastic collisions near the center of the galaxy (as well as near
each massive center) lead to the sharp heating of decelerated material
densities and to a local thermonuclear reaction for the synthesis of
light elements.
The thermonuclear energy release next to the center of a star also
heats its visible surface. This results in visible luminosity of
thermonuclear stars even at the edge of the galaxy, where other falling
bodies are cooling according to the Laue--Mosengeil transport of heat.

A small star that is on an elongated elliptical orbit around a massive
center should be heated twice per cycle due to sudden decelerations by
small circular satellites at minimum star approximations to this
center. In this case, rapid losses of translational kinetic energy
occur under explosive growth of internal heat energy and temperature.
Thus, the relativistic heat variable predicts a double burst of
luminosity of stars for each cycle along an elongated orbit. Upon
detection, such thermokinetic effects can be modeled for a numerical
verification of the internal energy variable in accordance with the
relativistic heat transfer law.

Recall again that radiation and absorption of electromagnetic waves
change the relativistic full energy of a probe body even in a constant
gravitational field. There are no steady mechanical charges in reality
and this was stated by Cartesian physics as early as in 1629 [13].
In closing, the Newton mass physics is to be replaced by tensor gravitation
and tensor mechanics for vector energy flows of  the time-varying internal heat. New non-dual physics of non-empty warm space [7] corresponds
 to original Cartesian ideas of vortex
matter-extensions. Internal relativistic heat controls the visible
spatial motion due to the introduced principle of equipartition of
kinetic energies. Cartesian physics of time-varying charges can change
the Newton world paradigm of localized particles in empty space and can
shed some light on how to warm the nonlocal Universe in advanced
theories of continuous flows of heat energy.\looseness=1

\bigskip
{{\bf Acknowledgments}. 
I am very grateful to Shpetim Nazarko for his ICNFP-2015 conference
poster and for useful discussions.
}
%
%
%


%

\begin{thebibliography}{}

\bibitem{1}
{Bulyzhenkov}, {I.E.}:
{Int. J. Theor. Phys.}
{47},
{1261}
{2008})

\bibitem{2}
{Kuhn}, {T.S.}:
{The Structure of Scientific Revolutions}.
{University of Chicago Press},
{Chicago}
({1962})


\bibitem{3}
{Landau}, {L.D.},
{Lifshitz}, {E.M.}:
{The Classical Theory of Fields},
{4}th edn.
{Course of Theoretical Physics Series},
vol.~{2}
({1980}).
{Butterworth-Heinemann}



\bibitem{4}
{Mosengeil}, {K.}:
{Ann. Phys. (Leipz.)}
{327},
{867}
({1907})

\bibitem{5}
{Umov}, {N.A.}:
Beweg-Gleich. d. Energie in contin. Korpern.
(Schomilch, Zeitschriff d. Math. und Phys., vol. XIX, 1874).
Selected works (in Russian), Izd. TTL (1950)
(Schomilch, Zeitschriff d. Math. und Phys., Bd. XIX,
1874); Selected Works (in Russian), Izd. TTL (1950)

\bibitem{6}
{Schwarzschild}, {K.}:
In: {Sitzungsber. Preuss. Akad. Wiss. Berlin (Math. Phys.)},
p.~{189}
({1916})

\bibitem{7}
{Einstein}, {A.},
{Infeld}, {L.}:
{The Evolution of Physics}.
{Cambridge University Press},
{Cambridge}
({1938})










\bibitem{8}
{Bulyzhenkov}, {I.E.},
{Pure Field Electrodynamics of Continuous Complex Charges}.
{MIPT (State University)},
{Moscow}
({2015}).
{4th year tutorial in Nonlinear Electrodynamics. ISBN
978-5-7417-0554-4.
Bull. Lebedev Phys. Inst., {43}, 138 (2016)}

\bibitem{9}
{Einstein}, {A.},
{Grossmann}, {M.}:
{Z. Math. Phys.}
{62},
{225}
({1913})

\bibitem{10}
{Einstein}, {A.}:
{Ann. Phys.}
{49},
{769}
({1916})

\bibitem{11}
{Einstein}, {A.}:
{Ann. Math.}
{40},
{922}
({1939})

\bibitem{12}
{Logunov}, {A.A.}:
{The Theory of Gravity}.
{Nauka},
{Moscow}
({2001})

\bibitem{13}
{Landau}, {L.D.},
{Lifshitz}, {E.M.}:
{Fluid Mechanics},
{2}nd edn.
{Course of Theoretical Physics Series},
vol.{6}.
{Pergamon Press},
{Oxford}
({1987})


\bibitem{G}
{Garber}, {D.}:
{Descartes' Metaphysical Physics}.
{University of Chicago Press},
{Chicago}
({1992})











\end{thebibliography}
\end{document}